\documentclass{elsart}
\usepackage{amssymb}
\begin{document}

\begin{frontmatter}

\title{\bf{Perturbations for the Coulomb - Kepler problem on de Sitter space-time }}
\author{Pop Adrian Alin}
\ead{apop302@quantum.physics.uvt.ro}
\address{{\small \it West University of Timi\c soara,} \\
 {\small \it V.  P\^ arvan Ave.  4, RO-300223 Timi\c soara, Romania}}

\begin{abstract}
In the Schr\" odinger picture of the Dirac quantum mechanics, defined in charts
with spatially flat Robertson-Walker metrics and Cartesian coordinates the perturbation theory is applied to the interacting part
of the Hamiltonian operator produced by the minimal coupling with the
gravitational field. First and second order perturbations are computed.
\end{abstract}
\end{frontmatter}
\maketitle
\section {Introduction}
	The purpose of this paper is to investigate the possible influences of the gravitational field over a system such as the hydrogen atom in a de Sitter space-time. Do such influences exist and if so, are these measurable by any experiment ?

	Recently a new time-evolution picture of the Dirac quantum mechanics was defined in charts with spatially flat Robertson-Walker metrics, under the name of Schr\" odinger picture \cite{coco}. In the non-relativistic quantum mechanics the time evolution can be studied in
different pictures (e. g., Schr\" odinger, Heisenberg, Interaction) which
transform among themselves through  specific time-dependent unitary
transformations. It is known that the form of the Hamiltonian operator and the
time dependence of other operators strongly depend on the picture choice. In
special and general relativity, despite of its importance, the problem of
time-evolution pictures is less studied because of the difficulties in finding
suitable Hamiltonian operators for scalar or vector fields.  

	We are interested in the Schr\" odinger picture defined in charts with spatially flat Robertson-Walker metrics and Cartesian coordinates. We take a look at the Schr\" odinger equation in this picture.  On a space-time with Robertson-Walker metric the Coulomb potential has the form $\frac{q^2}{r}\alpha(t)$. In the moving charts with RW metrics of the de Sitter spacetime the operator $H(t)$ is time-independent and $\dot{\alpha}/\alpha
=$const.  In this conditions the Coulomb potential in the Schr\" odinger picture is just $\frac{q^2}{r}$ so we find that the Schr\" odinger equation for the hydrogen atom is identical with the one from the non-relativistic quantum mechanics. This is the perfect scenario to apply the perturbation theory.

	Trying to add the gravitational interaction to the simple problem of the hydrogen atom gives us a very interesting case to study.
The Coulomb - Kepler problem is not solved directly so far by the use of the Dirac or Schr\" odinger equations. We propose to study the effects of the gravitational interaction by considering gravity as a perturbation added to the simpler problem of the hydrogen atom. For this paper we use numerical methods to compute the perturbations for the energy levels  produced by the minimal coupling with the gravitational field. We set the entire problem on the de Sitter space-time and apply the perturbation theory.

	We start in the second section with a presentation of the Schr\" odinger picture and by writing down the Schr\" odinger equation we will be using and an explanation of why it has this form  in our particular case. The next section is devoted to the perturbation theory and how we applied it in our case. A brief description of the numerical method used is also given as well as the results that were obtained with it. In the last section we are drawing a conclusion from these results.

\section{Preliminaries}
	We consider the Schr\" odinger picture as the picture in which the
kinetic part of the Dirac operator takes the standard form
$i\gamma^0\partial_t+i\gamma^i\partial_i$ \cite{coco}. The transformation
$\psi(x)\to \psi_S(x)=U_S(x)\psi(x)$ leading to the Schr\" odinger
picture is produced by the operator of time dependent {\em
dilatations}
\begin{equation}\label{U}
U_S(x)=\exp\left[-\ln(\alpha(t))(\vec{x}\cdot\vec{\partial})\right]\,,
\end{equation}
	The Schr\" odinger picture may offer one some technical
advantages in solving problems of quantum systems interacting with the
gravitational field. For example, in this picture we can derive the
non-relativistic limit (in the sense of special relativity) replacing ${\cal
H}_0$ directly by the Schr\" odinger kinetic term $\frac{1}{2m}{\vec{P}_S}^2$, where
the Hamiltonian operator ${\cal H}={\cal H}_0 + {\cal H}_{int}$.
Thus we obtain the Schr\" odinger equation
\begin{equation}\label{Sc}
\left[-\frac{1}{2m}\Delta -i\,
\frac{\dot{\alpha}(t)}{\alpha(t)}\left(\vec{x}\cdot \vec{\partial
}+\frac{3}{2}\right)\right]\phi(x)=i\partial_t \phi(x)\,,
\end{equation}
for the wave-function $\phi$ of a spinless particle of mass $m$.\\
In the particular case of the de Sitter spacetime, the  Schr\" odinger picture
will lead to important new results for the Schr\" odinger equation
in moving charts with RW metrics and spherical coordinates. In these charts
where $H=H(t)$ is conserved, the mentioned equation (\ref{Sc}) is analytically solvable in terms of Whittaker hypergeometric functions.

	We consider the interaction hamiltonian term ${\cal H}_{int}=\vec{x}\cdot \vec{\partial
}+\frac{3}{2}$ as a perturbation and would like to apply the well known methods of the perturbation theory to a simple case like the hydrogen atom. Our equation in physical units is
\begin{equation}\label{ScPh}
\left[-\frac{\hbar^2}{2m}\Delta -i\hbar\omega\left(\vec{x}\cdot \vec{\partial
}+\frac{3}{2}\right)+\frac{q^2}{r}\right]\Psi(x)=E \Psi(x)\,,
\end{equation}
where $q^2=\frac{Ze^2}{4\pi\epsilon_0}$ is the Coulomb potential (and $Z=1$ for the hydrogen atom). It can be easily shown that on de Sitter space-time the Coulomb potential has the same form ($\frac{q^2}{r}$) as on Minkowski space-time; in this way the "clasical" problem of the hydrogen atom remains unchanged and the perturbation theory is well suited for studying the effects of minimal coupling with the gravitational field.
\section{Perturbations - first and second order}
	It is obviously much better for such a case as the hydrogen atom to work with spherical coordinates. Making the transformation $\{t,\vec{x}\} \rightarrow \{t,r,\theta,\phi\}$ we get
$\vec r \vec \partial= r\frac{\partial}{\partial r}$. We also have the posibility now to separate the radial components from the spherical components:
\begin{equation}
<\vec x |n,l,m>=<r,\theta,\phi |n,l,m>;  \Psi(r,\theta,\phi)=R_{nl}(r)Y_{lm}(\theta ,\phi)
\end{equation}
	The radial functions are well known and can be expressed in terms of the Laguerre polynoms:
\begin{equation}
R_{nl}(r)=-\left\{\frac{2Z}{na_0}\frac{(n-l-1)!}{2n[(n+l)!]^3}\right\} ^{\frac{1}{2}}e^{(-\frac{\rho)}{2}}\rho ^lL_{n+l}^{2l+1}(\rho)
\end{equation}
where $\rho=\frac{2Z}{na}r$. The angular components of $\Psi(r,\theta,\phi)$ have to satisfy the condition $\int Y_{lm}^* Y_{l'm'}d\Omega = \delta_{ll'}\delta_{mm'}$ so we will only have to deal with the radial components from this point on. \\

	The energy levels in the case of the hydrogen atom are given by 
$E_{n}^{(0)}=\frac{me_{0}^{4}}{2\hbar ^2}\frac{1}{n^2}$; by adding corrections due to perturbations we have $$\Delta E = E_{n}^{(0)}+E_{n}^{(1)}+E_{n}^{(2)}+...$$. We want to compute this corrections for the first and second order.
For the first order we just need the matrix elements of the interacting hamiltonian term.
\begin{equation}
E_{n}^{(1)}=<n,l,m|H_{int}|n'l'm'>
\end{equation}
	The computation is straightforward and the result is zero for any value of n (energy level). This means the first order perturbations give no contribution to the correction of the energy levels.

	For the second order we used the well known relations from the perturbation theory that give us the second order perturbations of the energy levels.
\begin{equation}\label{pert2}
E_{n}^{(2)}=\sum_{k\neq n}\frac{<E_n|H_{int}|E_k><E_k|H_{int}|E_n>}{E_n - E_k}
\end{equation}
	This time te computation gets more complicated but still solvable with a mathematical software like Maple or Mathematica or by implementing a custom numerical routine. We define a procedure that can compute the values of the radial functions using the Laguerre polynoms by taking as input only the values of the quantum numbers $n$ and $l$. Using this we write the action of $H_{int}$ over $|E_k>$ and we can evaluate the integrals from the equation (\ref{pert2}). Than is just a matter of making the summation of the terms of this equation omitting  the term where $k\neq n$. By controling the number of terms in this summation we cand adjust the precision of our method.
The contributions due to the second order perturbations are not zero like those of the first order.
 
	We will give hear the results for the first values of $n$ that were obtained with some custom Maple procedures, with $Z=1$ for the hydrogen atom. 
 We have 
\begin{equation}\label{dimens}
E_{n}^{(2)}=\frac{\hbar^4\omega^2}{m e_0^4}\epsilon^{(2)}
\end{equation}
to get to a system of units with physical meaning.  Bellow  are given some of the computed values for the dimensionless quantity $\epsilon^{(2)}$.
 \begin{table}[ht!]
 \begin{center}
  \begin{tabular}{|c|c|c|c|c|c|c|}
\hline
\multicolumn{1}{|l|}{n} & l & \multicolumn{5}{c|}{Number of terms} \\ 
\cline{3-7}
\multicolumn{1}{|l|}{} &  & 10 & 30 & 50 & 100 & 1000 \\ 
\hline
1 & 0 & 1.064095523 & 1.073619436 & 1.074421747 &  1.074765599 & 1.074880495 \\ 
\hline
2 & 0 & 18.46605472 & 18.67877161  & 18.69583748 & 18.70311224 & 18.70553792 \\ 
\hline
2 & 1 & 13.70700332 & 13.83414505 & 13.84427053 & 13.84858330 & 13.85002081 \\ 
\hline
3 & 0 & 95.93893645 & 97.31876631 & 97.42041641 & 97.46335870 & 97.47762648 \\ 
\hline
3 & 1 & 84.69291072 & 85.74907706 & 85.82617976 & 85.85872010 & 85.86952714 \\ 
\hline
3 & 2 & 60.91483176 & 61.40294550 & 61.43776962 & 61.45243081 & 61.45729539 \\ 
\hline
4 & 0 & 305.9528653 & 311.5523974 & 311.9151219 & 312.0664192 & 312.1164354 \\ 
\hline
4 & 1 & 285.5680992 & 290.2868876 & 290.5892387 & 290.7152100 & 290.7568349 \\ 
\hline
4 & 2 & 243.4746022 & 246.5590712 & 246.7515894  & 246.8315844 & 246.8579881 \\ 
\hline
4 & 3 & 177.1352766 & 178.3299632 & 178.4009022 & 178.4302308 & 178.4398912 \\ 
\hline
\end{tabular}
\end{center}

\caption {\it{Perturbation for the energy levels $\epsilon_{n}^{(2)}$ computed with different number of terms}}

\label{tab:Tabel}
\end{table}

	For the equation (\ref{pert2}) the same  numerical routines were applied making the sumation for different number of terms, from $10$ to $1000$. This was a test of the convergence of our method of computation; the results did not  change significantly even though  the number of terms was greatly increased.

	We ask ourselves what is the magnitude of this perturbation in order to see if it can be measured or not, and if so what to expect. For this we evaluate from the equation (\ref{dimens}) the factor that multiplies $\epsilon^{(2)}$ by having the value of $\omega$ of the same order as the Hubble constant ($H_0 \approx 2.3 \times 10^{-18}s^{-1}$) in our de Sitter universe. This gives us : $\frac{\hbar^4\omega^2}{m e_0^4}=\frac{(\hbar\omega)^2}{1 u.a.e}\approx 8.4 \times 10^{-68} eV$ ($1 u.a.e. = 27.21 eV$). This value is extremely small and obviously can not be measured in any experiment. 
\section{Concluding remarks}
	Let us summarize what was done: we have started with the classical hydrogen atom problem by taking advantage of the fact that  on de Sitter space-time the Coulomb potential has the same form as on Minkowski space-time. We have considered the gravitational interaction as a perturbation and using the perturbation theory we have computed the first and second order perturbations for the energy levels of the atom. In the first order we have seen that all this are zero. In the second order although we got non-zero values those are so small that can easily be neglected and can not be even measured in any kind of experiment.
The conclusion we draw is that theere are no measurable cosmological influences over the atoms from the minimal coupling with the gravitational field.\\

\textbf{Acknowledgements}
  \par
	We would like to thank Professor Ion I.Cot\u aescu  for encouraging us to do this work and for reading the manuscript.

\end{document}